\documentstyle[preprint,aps,epsfig]{revtex}

\begin{document}
\title{Coherent control of atom dynamics in an optical lattice}
\author{H.L. Haroutyunyan and G. Nienhuis\thanks{}}
\address{Huygens Laborotarium, Universiteit Leiden,\\
Postbus 9504, \\
2300 RA Leiden, The Netherlands}
\maketitle

\begin{abstract}
On the basis of a simple exactly solvable model we discuss the possibilities
for state preparation and state control of atoms in a periodic optical
potential. In addition to the periodic potential a uniform force with an
arbitrary time dependence is applied. The method is based on a formal
expression for the full evolution operator in the tight-binding limit. This
allows us to describe the dynamics in terms of operator algebra, rather than
in analytical expansions.
\end{abstract}

\section{Introduction}

The energy eigenvalues of a quantum particle moving in a periodic potential
form energy bands (the Bloch bands) that are separated by bandgaps. The
eigenstate within a band is characterized by the quasimomentum, which
determines the phase difference between two points separated by a period. An
initially localized wavepacket typically propagates through space, leading
to unbounded motion. When an additional uniform force is applied, the Bloch
bands break up into a ladder of equally spaced energy levels, which are
called the Wannier-Stark ladder. In this case, a wavepacket of the particle
extending over several periods can exhibit bounded oscillatory motion,
termed Bloch oscillation, at a frequency determined by the level separation
in the ladder. These early results of the quantum theory of electrons in
solid crystals \cite{BLO29,LAN32,ZEN34,WAN60} have regained interest
recently due to the advent of optical lattices for atoms. These lattices are
formed when cold atoms are trapped in the periodic potential created by the
superposition of a number of traveling light waves \cite
{VER92,JES92,HEM93,JES96}. In contrast to the case of electrons in crystal
lattices, these optical lattice fields have virtually no defects, they can
be switched on and off at will, and dissipative effects can be largely
controlled. The phenomenon of Bloch oscillations has first been observed for
cesium atoms in optical lattices \cite{BEN96}. The uniform external force is
mimicked by a linear variation of the frequency of one of the
counterpropagating traveling waves, thereby creating an accelerated standing
wave. By applying a modulation on the standing-wave position, Rabi
oscillations between Bloch bands, as well as the level structure of the
Wannier-Stark ladder has been observed for sodium atoms in an optical
lattice \cite{WIL96}. Theoretical studies of transitions between ladders
have also been presented \cite{GLU00}. Bloch oscillations have also been
demonstrated for a light beam propagating in an array of waveguides, with a
linear variation of the refractive index imposed by a temperature gradient 
\cite{PER99}.

When the applied uniform force is oscillating in time, the motion of a
particle in a periodic potential is usually unbounded. However, it has been
predicted that the motion remains bounded for specific values of the ratio
of the modulation frequency and the strength of the force \cite{DUN86}.
Similar effects of dynamical localization, including routes to chaos, have
been studied experimentally for optical lattices, including both amplitude
and phase modulation of the uniform force \cite{BHA99}. Phase transitions
have been predicted for atoms in two incompatible periodic optical
potentials imposed by bichromatic standing light waves \cite{DRE97}.

In the present paper we discuss the Wannier-Stark system with a
time-dependent force, as a means of preparing the state of particles in a
periodic potential. We derive an exact expression for the evolution operator
of the particle, with an arbitrary time-dependent force. This allows one to
apply the combination of delocalizing dynamics in the absence of the uniform
force with the periodic dynamics induced by a uniform force for coherent
control of the state of the particles. Exact solutions in the case of a
constant uniform force have been obtained before by analytical techniques 
\cite{LUB85,DAT87}. The operator method phenomena induced by an oscillating
force to be described exactly in a unified scheme. Examples are dynamical
localization and fractional Wannier-Stark ladders.

The model is described in one dimension. However, this is no real
restriction. Under the assumption of nearest-neighbor interaction, the
corresponding 2D or 3D problem exactly factorizes into a product of 1D
solutions.

\section{Model system}

\subsection{Periodic potential}

\label{perpot} The quantummechanical motion of atoms in a periodic optical
potential $V(x)$ with period $a$, is described by the Hamiltonian 
\begin{equation}
H_0 = {\frac{P^2 }{2M}} + V(x)\; .  \label{H0}
\end{equation}
We assume that the atoms are sufficiently cooled, so that only the lowest
energy band is populated. The ground state in well $n$ located at $x = na$
is indicated as $|n\rangle $. These states play the role of the basis of
localized Wannier states. For simplicity we make the tight-binding limit,
where only the ground levels in neighboring wells are coupled. When we
choose the zero of energy at the ground level in a well, the Hamiltonian (%
\ref{H0}) projected on these ground levels is defined by 
\begin{equation}
H_0 = {\frac{1 }{2}} \hbar \Omega(B_+ + B_-)\;\;, \;\;B_{\pm}|n\rangle =
|n\pm1\rangle\;.  \label{H0disc}
\end{equation}
The raising and lowering operators $B_+$ and $B_-$ are each other's
Hermitian conjugate, and each one of them is unitary. The frequency $\Omega$
measures the coupling between neighboring wells, due to tunneling through
the barriers. We shall allow the coupling to depend on time. The eigenstates
of $H_0$ are directly found by diagonalizing the corresponding matrix. These
states are the Bloch states $|k\rangle$, with energy $E(k) = \hbar \Omega
\cos(ka)$. Their expansion in the Wannier states, and the inverse relations
can be expressed as 
\begin{equation}
|k\rangle = \sqrt{\frac{a }{2\pi}} \sum_n e^{inka} |n\rangle \;\;,\;\;
|n\rangle = \sqrt{\frac{a }{2\pi}} \int\; dk\; e^{-inka} |k\rangle \;\;.
\label{bloch}
\end{equation}
Obviously, the states $|k\rangle$ are periodic with period $2 \pi/a$, and
the quasimomentum $k$ can be chosen from the Brillouin zone $[-\pi/a,\pi/a]$%
. The integration in (\ref{bloch}) extends over this Brillouin zone. From
the translation property $\langle x|n\rangle = \langle x+a|n+1\rangle$ of
the Wannier wave functions it follows that the states (\ref{bloch}) do
indeed obey the Bloch condition $\langle x+a|k\rangle = \exp(ika) \langle
x|a\rangle$. When the states $|n\rangle $ are normalized as $\langle
n|m\rangle = \delta_{nm}$, the Bloch states obey the continuous
normalization relation $\langle k|k^{\prime}\rangle = \delta(k - k^{\prime})$%
.

\subsection{Uniform force}

\label{uniform} An additional uniform force is described by adding to the
Hamiltonian the term 
\begin{equation}
H_1 = {\frac{\hbar x \Delta }{a}}\;,  \label{H1}
\end{equation}
where the (possibly time-dependent) force of size $\hbar \Delta(t)/a$ is in
the negative direction. On the basis of the Wannier states, this term is
diagonal, and it is represented as 
\begin{equation}
H_1 = \hbar \Delta B_0\;\;,\;\; B_0|n\rangle = n |n\rangle\;\;.
\label{H1disc}
\end{equation}
Hence the evolution of a particle under the influence of the total
Hamiltonian 
\begin{equation}
H = H_0 + H_1\;,  \label{Hdisc}
\end{equation}
with $H_0$ and $H_1$ defined by eqs. (\ref{H0disc}) and (\ref{H1disc}), in
terms of the operators $B_{\pm}$ and $B_0$. We shall also need expressions
for the operators $B_{\pm}$ and $B_0$ acting on a Bloch state. These can be
found from the definition of the operators and the expansions (\ref{bloch}).
One easily finds that 
\begin{equation}
B_{\pm} |k \rangle = e^{\mp ika} |k \rangle\;,\; e^{-i \beta B_0} |k \rangle
= |k - {\frac{\beta }{a}} \rangle\;.  \label{Bbloch}
\end{equation}
In Bloch representation the operators have the significance $B_{\pm} =
\exp(\mp ika)$, $B_0 = (i/a)(d/dk)$, which is confirmed by the commutation
rules (\ref{comm}). The Wannier states may be viewed as discrete position
eigenstates, with $B_0$ the corresponding position operator. The Bloch
states play the role of momentum eigenstates, and the finite range of their
eigenvalues within the Brillouin zone reflects the discreteness of the
position eigenvalues.

\subsection{Operator algebra}

The basic operators $B_{\pm}$ and $B_0$ obey the commutation rules 
\begin{equation}
[B_0, B_{\pm}] = \pm B_{\pm}\;,\; [B_+,B_-] = 0\;.  \label{comm}
\end{equation}
In order to derive exact expressions for the evolution operator
corresponding to the Hamiltonian (\ref{Hdisc}), we need several operator
identities involving these operators $B_0$ and $B_{\pm}$. The identities 
\begin{equation}
e^{i \beta B_0} B_{\pm} e^{-i \beta B_0} = e^{\pm i \beta} B_{\pm}\;.
\label{transpm}
\end{equation}
directly follow from the commutation rules (\ref{comm}), and they lead the
transformation rules 
\begin{equation}
e^{i \beta B_0} \exp\Big( -i {\frac{1 }{2}} \alpha(B_+ + B_-) \Big) e^{-i
\beta B_0} = \exp\Big(-i {\frac{1 }{2}} \alpha ( e^{i \beta} B_+ + e^{-i
\beta} B_- )\Big)  \label{trans}
\end{equation}
for arbitrary values of $\alpha$ and $\beta$. We shall also need the
equalities 
\begin{equation}
\exp({\frac{i }{2}}\alpha B_{\pm}) B_{0} \exp(-{\frac{i }{2}} \alpha
B_{\pm}) = B_0 \mp {\frac{i }{2}} \alpha B_{\pm}\;,  \label{trans0}
\end{equation}
which are verified after differentiation with respect to $\alpha$, while
using the commutation rules (\ref{comm}).

\section{Operator description of evolution}

\subsection{Evolution operator}

\label{evolution} In this section we derive expressions for the evolution
operator $U(t,0)$, which transforms an arbitrary initial state $|\Psi(0)
\rangle$ as $|\Psi(t) \rangle = U(t,0) |\Psi(0) \rangle$. The results are
valid for any time-dependence of the uniform force and the coupling between
neighboring wells, as specified by $\Delta(t)$ and $\Omega(t)$. A
time-dependent coupling represents the case that the intensity of the
lattice beams is varied. We express the evolution operator in the factorized
form 
\begin{equation}
U(t,0) = U_1(t,0) U_0(t,0)\;,  \label{Ufact}
\end{equation}
where $U_1(t,0) = \exp[-i \phi(t) B_0]$ gives the evolution corresponding to
the Hamiltonian $H_1$ alone, in terms of the phase shift 
\begin{equation}
\phi(t) = \int_0^t\; dt^{\prime}\; \Delta(t^{\prime})\;.  \label{phi}
\end{equation}
From the evolution equation for $U$ with the Hamiltonian (\ref{Hdisc}),
while using the transformation (\ref{transpm}) we find the evolution
equation 
\begin{equation}
{\frac{dU_0 }{dt}} = - {\frac{i \Omega(t) }{2}} \Big( e^{i \phi(t)} B_+ +
e^{-i \phi(t)} B_-\Big) U_0(t)\;.  \label{U_0}
\end{equation}
Since this equation only contains the commuting operators $B_+$ and $B_-$,
it can easily be integrated. In fact, the solution is given by eq. (\ref
{trans}) with the time-dependent values of the real parameters $\alpha$ and $%
\beta$ defined by the relations 
\begin{equation}
\alpha(t) e^{i \beta(t)} = \int^t_0\; dt^{\prime}\Omega(t^{\prime}) e^{i
\phi(t^{\prime})}\;.  \label{alphabeta}
\end{equation}
Combining this solution with the definition of $U_1$, leads to a closed
expression for the evolution operator $U(t,0)$ for an arbitrary time
dependence of the uniform force, in terms of the parameters $\alpha$, $\beta$
and $\phi$, defined in (\ref{phi}) and (\ref{alphabeta}). The result is $%
U(t,0) \equiv R(\alpha, \beta, \phi)$, with $R$ defined by 
\begin{equation}
R(\alpha, \beta, \phi) = e^{i(\beta - \phi)B_0} \exp\Big( -i {\frac{1 }{2}}
\alpha(B_+ + B_-) \Big) e^{-i \beta B_0}\;.  \label{Ut}
\end{equation}
This defines the unitary operator $R$ as a function of the three parameters $%
\alpha$, $\beta$ and $\phi$. The result is valid for an arbitrary time
dependence of the force and the coupling, described by $\Delta(t)$ and $%
\Omega(t)$. The characteristics of the evolution of an arbitrary initial
state is determined by the properties of the operators $R$ as a function of $%
\alpha$, $\beta$ and $\phi$. Mathematically, these operators form a
three-parameter group, which is generated by the three operators $B_{\pm}$
and $B_0$.

On the basis of the Wannier states, the contribution of the operator $B_0$
in (\ref{Ut}) is trivial, whereas the effect of the exponent containing $%
B_{\pm}$ can be evaluated by first expanding a Wannier state in Bloch
states, for which the action of this exponent is simple. Then reexpressing
the Bloch states in Wannier states, we find 
\begin{equation}
\exp\Big( -i {\frac{1 }{2}} \alpha(B_+ + B_-) \Big) |m\rangle = \sum_n
i^{-n+m} J_{n-m}(\alpha) |n\rangle\;,  \label{Bwan}
\end{equation}
where we used the defining expansion $\exp(i \xi \sin \phi) = \sum_n
\exp(in\phi) J_n(\xi)$ of the ordinary Bessel functions. Hence the matrix
elements of the operator (\ref{Ut}) between Wannier states are 
\begin{equation}
\langle n| R(\alpha, \beta, \phi) |m \rangle = (i e^{-i \beta})^{-n+m}
e^{-in\phi} J_{n-m}(\alpha) \;.  \label{Uwan}
\end{equation}

For the evolution operator (\ref{Ut}) in Bloch representation we can just
use the form of the operators $B_{\pm}$ and $B_0$, as given in Sec. \ref
{uniform}. This leads to the result 
\begin{equation}
R(\alpha, \beta, \phi) |k\rangle = e^{-i \alpha \cos(ka - \beta)} |k -
\phi/a \rangle\;.  \label{Ubloch}
\end{equation}
This shows that the quasimomentum as a function of time varies as $k(t) =
k(0) - \phi(t)/a$, with $\phi(t)$ given in (\ref{phi}). The parameter $\phi$
determines the shift of the quasimomentum during the evolution. The
expressions (\ref{Uwan}) and (\ref{Ubloch}) clarify the significance of the
three parameters $\alpha$, $\beta$ and $\phi$ that specify the evolution
operator.

\subsection{Heisenberg picture}

\label{heis} The transport properties of any initial state is conveniently
described by the evolution of the operators in the Heisenberg picture. Since
any evolution operator can be written in the form of $R(\alpha, \beta, \phi)$
for the appropriate values of the parameters, we can view $R^{\dagger} B R$
as the Heisenberg operator corresponding to any operator $B$. The Heisenberg
operators corresponding to $B_{\pm}$ can be expressed as 
\begin{equation}
R^{\dagger} (\alpha, \beta, \phi) B_{\pm} R(\alpha, \beta, \phi) = e^{\pm i
\phi} B_{\pm} \;,  \label{heisBpm}
\end{equation}
which is directly shown by using eq. (\ref{transpm}). Since $B_{\pm} = \exp({%
\mp}ika)$ in Bloch representation, this confirms the significance of $\phi$
as the shift of the value of the quasimomentum.

After using the transformation property (\ref{trans0}), one finds the
Heisenberg operator corresponding to the position operator $B_0$ as 
\begin{equation}
R^{\dagger} (\alpha, \beta, \phi) B_0 R(\alpha, \beta, \phi) = B_0 + {\frac{%
i \alpha }{2}}(e^{-i \beta} B_- - e^{i \beta} B_+) \;.  \label{heisB0}
\end{equation}
This implies that the expectation value of the position after evolution is
determined by 
\begin{equation}
\langle n \rangle = \langle B_0 \rangle + {\frac{i \alpha }{2}}(e^{-i \beta}
\langle B_- \rangle - e^{i \beta} \langle B_+ \rangle)\;,  \label{<n>}
\end{equation}
where the averages in the r.h.s. should be taken with respect to the inital
state. Hence no displacement of a wavepacket can occur whenever $\langle B_+
\rangle = \langle B_- \rangle ^* = 0$. This is true whenever the initial
state is diagonal in the Wannier states $|n\rangle$. Conversely, average
motion of a wavepacket can only occur in the presence of initial phase
coherence between neigboring Wannier states. The width of a wavepacket is
determined by the expectation value of the square of the Heisenberg position
operator (\ref{heisB0}). This gives the expression 
\begin{eqnarray}
\langle n^2 \rangle &=& \langle B_0^2 \rangle + {\frac{\alpha^2 }{4}} (2 -
e^{-2 i \beta} \langle B_-^2 \rangle - e^{2 i \beta} \langle B_+^2 \rangle) 
\nonumber \\
&+& {\frac{i \alpha }{2}}(e^{-i \beta} \langle B_0 B_- + B_- B_0 \rangle -
e^{i \beta} \langle B_0 B_+ + B_+ B_0 \rangle) \;,  \label{<n2>}
\end{eqnarray}

\section{Localized initial states}

\label{specinit}

\subsection{Arbitrary wavepackets}

A fairly localized initial state $|\Psi(0)\rangle = \sum_n c_n |n\rangle$
with a reasonably well-defined quasimomentum can be modelled by assuming
that neigboring states have a fixed phase difference $\theta$, so that 
\begin{equation}
c_{n}^* c_{n+1} = |c_{n} c_{n+1}| e^{i \theta}\;.  \label{momcond}
\end{equation}
Then the quasimomentum is initially centered around the value $k_0 =
\theta/a $. For simplicity, we assume moreover that the distribution over
Wannier states is even in $n$, so that $|c_n| = |c_{-n}|$. The initial
average position of the particle is located at $n=0$. In order to evaluate
the time-dependent average position and spreading of the packet, we can
apply eqs. (\ref{<n>}) and (\ref{<n2>}). The symmetry of the distribution
implies that $\langle B_0 \rangle = 0$, while $\langle B_0^2 \rangle =
\sigma_0^2$ is the initial variance of the position. When we introduce the
quantities 
\begin{equation}
\sum_n |c_{n+1} c_n| \equiv b_1\;,\; \sum_n |c_{n+2} c_n| \equiv b_2\;,
\label{offdiag}
\end{equation}
we obtain the simple identities 
\begin{equation}
\langle B_+ \rangle = b_1 e^{-i \theta}\;,\; \langle B_+^2 \rangle = b_2
e^{-2 i \theta}\;,\; \langle B_0 B_+ \rangle = - \langle B_+ B_0 \rangle = {%
\frac{1 }{2}} b_1 e^{-i \theta}\;.  \label{expval}
\end{equation}
The last identity is proven by using that the quantity $f_{2n+1} \equiv
|c_{n+1} c_n|$ is even in its index (which takes only odd values).
Therefore, $\sum_l l f_l = 0$, which is equivalent to the statement that $%
2\langle B_+ B_0 \rangle + \langle B_+ \rangle = 0$. The other expectation
values occurring in eqs. (\ref{<n>}) and (\ref{<n2>}) are found by taking
the complex conjugates of the identities (\ref{expval}). This leads to the
simple exact results 
\begin{equation}
\langle n \rangle = \alpha b_1 \sin(\beta - \theta) \;\;,\;\; \langle n^2
\rangle = \sigma_0^2 + {\frac{\alpha^2 }{2}}\Big( 1 - b_2 \cos 2(\beta -
\theta) \Big)\;,  \label{<n2>spec}
\end{equation}
so that the variance of the position is found as 
\begin{equation}
\sigma^2 \equiv \langle n^2 \rangle - \langle n \rangle ^2 = \sigma_0^2 + {%
\frac{\alpha^2 }{2}} \Big( 1 - b_1^2 - (b_2 - b_1^2) \cos 2(\beta - \theta) %
\Big)\;.  \label{var}
\end{equation}

Notice that the parameters $b_1$ and $b_2$ are real numbers between $0$ and $%
1$. In the limit of a wide initial wavepacket, determined by coefficients $%
c_n$ whose absolute values vary slowly with $n$, the parameters $b_1$ and $%
b_2$ both will approach $1$, and the width $\sigma$ will not vary during the
evolution. In the opposite special case that the initial state is the single
Wannier state $|0 \rangle$, one finds that $b_1 = b_2 = 0$, so that the
width $\sigma = \alpha/\sqrt {2}$.

In the special case that the particle is initially localized in the single
Wannier state at $x = 0$, so that $|\Psi(0)\rangle = |0\rangle$, the
parameters $b_1$, $b_2$ and $\sigma_0$ vanish, so that 
\begin{equation}
\langle n \rangle = 0 \;,\; \sigma^2 = \langle n^2\rangle = \alpha^2/2\;.
\label{msd}
\end{equation}
This shows that the average position of the wavepacket does not change, and
that its width is determined by the parameter $\alpha$ alone. This is in
line with the fact that the population distribution over the Wannier states
after the evolution is $p_n = |\langle n |R|0\rangle |^2 = J_n^2(\alpha)$,
as follows from Eq. (\ref{Uwan}). Hence the (time-dependent) value of $%
\alpha $ determines the spreading of an initially localized particle.

\subsection{Gaussian wavepacket}

\label{Gaussian} When the initial distribution over the sites is Gaussian
with a large width, we can evaluate the full wavepacket after evolution.
Suppose that the initial state is specified by the coefficients 
\begin{equation}
c_n = {\frac{1 }{\sqrt {\sigma_0 \sqrt {2 \pi}}}} e^{i n \theta} \exp( - {%
\frac{n^2 }{4 \sigma_0^2}})\;,  \label{cinitial}
\end{equation}
which obey the condition (\ref{momcond}). This state is properly normalized
provided that $\sigma_0 \gg 1$. When the evolution operator is expressed as
in (\ref{Ut}), the time-dependent state is expanded as $|\Psi(t) \rangle =
R|\Psi(0)\rangle = \sum_n f_n \exp [in(\theta - \phi)] |n \rangle$.
Summation expressions for the coefficients $f_n$ are directly obtained by
using the expression (\ref{Uwan}) of $R$ in Wannier representation. We use
similar techniques as applied in Ref. \cite{ISH00} in the context of the
diffraction of a Gaussian momentum distribution of atoms by a standing light
wave. The technique is based on differentation of the expression for $f_n$
with respect to $n$, while using the property $\alpha[J_{n+1}(\alpha) +
J_{n-1}(\alpha)] = 2 n J_n(\alpha)$ of Bessel functions. When the width is
sufficiently large, so that the difference $f_{n+1} - f_n$ can be
approximated by the derivative, this leads to the differential equation 
\begin{equation}
2 \sigma_0^2 {\frac{d f_n }{dn}} \approx \Big(\alpha \sin(\beta - \theta) - n%
\Big) f_n + i \alpha \cos (\beta - \theta) {\frac{d f_n }{dn}}\;.
\label{diffeq}
\end{equation}
By solving this equation, we arrive at the closed expression 
\begin{equation}
f_n = {\frac{1 }{{\cal N}}}\; \exp \Big( {\frac{- n^2/2 + \alpha n \sin
(\beta - \theta) }{2 \sigma_0^2 - i \alpha \cos (\beta - \theta)}} \Big)\;,
\label{ffinal}
\end{equation}
with the normalization constant determined by 
\begin{equation}
{\cal N}^4 = \pi \Big(2 \sigma_0^2 + {\frac{\alpha^2 \cos^2(\beta - \theta) 
}{2 \sigma_0^2}}\Big)\;.  \label{norm}
\end{equation}

We find that the distribution is Gaussian at all times, with a time-varying
average position and variance. These are given by the expressions 
\begin{equation}
\langle n \rangle = \alpha \sin (\beta - \theta)\;,\; \sigma^2 = \sigma_0^2
+ {\frac{\alpha^2 }{8 \sigma_0^2}} \Big(1 + \cos 2(\beta - \theta) \Big)\;.
\label{<n>Gauss}
\end{equation}
These results are in accordance with eqs. (\ref{<n2>spec}) and (\ref{var}),
as one checks by using the approximate expressions $b_l = \exp(- l^2 / 8
\sigma_0^2) \approx 1 - l^2 / 8 \sigma_0^2$, while neglecting terms of order 
$(1/\sigma_0)^4$ and higher. The width of the packet never gets smaller than
its initial value. The phase difference between neighboring sites is mainly
determined by $\theta - \phi$. This shows that a phase difference can be
created or modified in a controlled way, simply by imposing a time-dependent
force that gives rise to the right value of $\phi$. Notice that in these
expressions (\ref{<n>Gauss}), $\theta$ and $\beta$ enter in an equivalent
fashion. The position and the width of the Gaussian distribution can be
controlled at will by adapting the force to the desired value of $\beta$.

We recall that the results of this chapter are valid for an arbitray
time-dependent force $\Delta (t)$, which determines the time-dependent
values of the parameters $\alpha$, $\beta$ and $\phi$ as specified in eqs. (%
\ref{phi}) and (\ref{alphabeta}). In the subsequent sections, we specialize
these expressions for constant or oscillating values of the uniform force.

\section{Constant uniform force and Bloch oscillations}

\subsection{Wannier-Stark ladder of states}

The case of a constant force is the standard situation where Bloch
oscillations occur. When $\Delta$ and $\Omega$ are constant, the Hamiltonian
is time-independent, and then it is convenient to introduce the normalized
eigenstates $|\psi_m \rangle$ of $H$. When we expand these eigenstates in
the Wannier states as $|\psi_m \rangle = \sum_n |n\rangle c_n^{(m)}$, the
eigenvalue relation $H|\psi_m \rangle = E_m |\psi_m \rangle$ with $E_m =
\hbar \omega_m$ leads to the recurrence relations for the coefficients 
\begin{equation}
{\frac{1 }{2}} \Omega \Big(c_{n-1}^{(m)} + c_{n+1}^{(m)}\Big) + \Delta n
c_n^{(m)} = \omega_m c_n^{m}\;.  \label{eigen}
\end{equation}
We introduce the generating function 
\begin{equation}
Z_m(k) = \sqrt{\frac{a }{2\pi}} \sum_n c_n^{(m)} e^{-inka}\;,  \label{Z}
\end{equation}
which is normalized for integration over the first Brillouin zone. In fact,
from the expression (\ref{bloch}) of the Bloch state, one notices that the
generating function $Z_m(k) = \langle k | \psi_m \rangle$ is equal to the
Bloch representation of the eigenstate $|\psi_m \rangle$. The relations (\ref
{eigen}) are found to be equivalent to the differential equation 
\begin{equation}
\Omega \cos (ka) Z_m(k) - {\frac{\Delta }{ia}} {\frac{d }{dk}} Z_m(k) =
\omega_m Z_m(k)\;,  \label{eigenrel}
\end{equation}
with the obvious normalized solution 
\begin{equation}
Z_m(k) = \sqrt{\frac{a }{2\pi}} \exp\Big({\frac{i }{\Delta}}[\Omega \sin(ka)
- ak \omega_m ] \Big)\;.  \label{eigenbloch}
\end{equation}
Since the functions $Z_m(k)$ as defined by (\ref{Z}) are periodic in $k$
with period $2 \pi/a$, the same must be true for the expressions (\ref
{eigenbloch}). Hence, the frequency eigenvalues must be an integer multiple
of $\Delta$, so that we can choose $\omega_m = m \Delta$, with integer $m$.
For these values of the eigenfrequencies, the coefficients $c_n^{(m)}$
follow from the Fourier expansion of $Z_m$, with the result 
\begin{equation}
c_n^{(m)} \equiv \langle n|\psi_m \rangle = J_{m - n}(\Omega/\Delta)\;.
\label{eigenwan}
\end{equation}

We find that the total Hamiltonian $H$ has the same eigenvalues as $H_1$.
Apparently, the energy shifts due to the coupling between the Wannier states
as expressed by $H_0$ cancel each other. Since the energy eigenvalues are
integer multiples of $\Delta$, each solution of the Schr\"odinger equation
is periodic in time with period $2 \pi/\Delta$, and the same is true for the
evolution operator $U(t)$ given in eq. (\ref{Ut}). This also implies that an
initial localized state remains localized at all times, due to the addition
of the uniform external force. The eigenstates $|\psi_m \rangle$ are the
Wannier-Stark ladder of states \cite{WIL96}. They form a discrete
orthonormal basis of the first energy band, and they are intermediate
between the Wannier and the Bloch basis of states.

\subsection{Oscillations of localized states}

The definitions (\ref{phi}) and (\ref{alphabeta}) show that 
\begin{equation}
\alpha = (2\Omega/\Delta) \sin (\Delta t/2)\;,\; \beta = \Delta t
/2\;,\;\phi = \Delta t\;.  \label{defparcst}
\end{equation}
In the Wannier representation, the matrix elements of $U$ are found from (%
\ref{Ut}) as 
\begin{equation}
\langle n| U(t,0) |m \rangle = i^{-n+m} e^{-i\Delta t (n + m)/2} J_{n-m}({%
\frac{2 \Omega }{\Delta}} \sin {\frac{\Delta t }{2}}) \;,  \label{Uwancst}
\end{equation}
which represents the transition amplitude from an initial state $|m \rangle$
to the final state $|n \rangle$. For the initial Wannier state $%
|\Psi(0)\rangle = |0\rangle$, the time-dependent state is $|\Psi(t)> =
\sum_n f_n(t) |n \rangle$ with 
\begin{equation}
f_n(t) = i^{-n} e^{-i \Delta t n/2} J_n({\frac{2 \Omega }{\Delta}} \sin {%
\frac{\Delta t }{2}})\;.  \label{fn}
\end{equation}
This is in accordance with Eq. (50) of ref. \cite{LUB85}, which has been
obtained by a rather elaborate analytical method, rather than an algebraic
one. Equation (\ref{msd}) shows that the time-dependent average position $%
\langle n \rangle$ of the wavepacket remains zero at all times, whereas the
mean-square displacement $\sigma = |\alpha|/\sqrt{2}$ displays a breathing
behavior, and returns to zero after the Bloch period $2 \pi / \Delta$.
Moreover, according to eq. (\ref{fn}), the phase difference between
neighboring sites varies continuously with time.

This is already quite different when only two Wannier states are populated
initially. Consider the initial state 
\begin{equation}
|\Psi(0)\rangle = {\frac{1 }{\sqrt{2}}} (|0\rangle + e^{i \theta} |1
\rangle)\;.  \label{twostate}
\end{equation}
Then the average position can be evaluated from eq. (\ref{<n>}), for the
values of $\alpha$ and $\beta$ as given in (\ref{defparcst}). The result is 
\begin{equation}
\langle n \rangle = {\frac{1 }{2}} + {\frac{\Omega }{2\Delta}} \Big( \cos
\theta - \cos (\Delta t - \theta) \Big)\;,  \label{<n>t2}
\end{equation}
which shows that the packet displays a harmonically oscillating behavior.
The amplitude of the oscillation is governed by the ratio $\Omega/\Delta$,
which is half the maximum amplitude for Bloch oscillations of a wavepacket
with a large width (see Sec. \ref{Gaussiant}). This amplitude must be
appreciable in order that interband coupling induced by the uniform force
remains negligible, as we have assumed throughout this paper. The
distribution $p_n = |f_n|^2$ after half a Bloch period, both for the initial
single Wannier state and for the inital state (\ref{twostate}) is
illustrated in Fig. 1. This demonstrates that a strong displacement can
already be induced by evolution of a superposition state of just two
neighboring Wannier states, with a specific phase difference. This
displacement arises from the interference between the transition amplitudes
from the two initial states to the same final state $|n \rangle$.

\subsection{Bloch oscillations and breathing of a Gaussian wavepacket}

\label{Gaussiant} The evolution of a Gaussian wavepacket as discussed in
Sec. {\ref{Gaussian} is specialized to the present case of a constant force
after substituting the expressions (\ref{defparcst}) in eqs. (\ref{ffinal})-(%
\ref{<n>Gauss}). We find for the average position $\langle n \rangle$ the
identity 
\begin{equation}
\langle n(t) \rangle = {\frac{\Omega }{\Delta}} [\cos \theta - \cos(\theta -
\Delta t)]\;.  \label{<n>t}
\end{equation}
This demonstrates that the wavepacket oscillates harmonically in position
with frequency $\Delta$, and with amplitude $\Omega/\Delta$ in units of the
lattice distance $a$. The velocity of the wavepacket is found from the time
derivative of (\ref{<n>t}), with the result 
\begin{equation}
v(t) = - a \Omega \sin(\theta - \Delta t)\;.  \label{v}
\end{equation}
It is noteworthy that this expression (\ref{v}) coincides exactly with the
expression for the group velocity $dE/\hbar dk$, with the derivative
evaluated at the time-dependent value of the quasimomentum $(\theta- \Delta
t)/a$, with $E = \hbar \Omega \cos(ka)$ the dispersion relation between
energy and quasimomentum in the absence of the uniform force, as given in
Sec. \ref{perpot}. Apparently, the expression for the group velocity retains
its validity also in the presence of the uniform force. Of course, the
concept of Bloch oscillations of the wavepacket as a whole has significance
only when the amplitude $\Omega /\Delta$ of the oscillation is large
compared with the width $\sigma$ of the packet, which in turn must extend
over many lattice sites. }

The time-dependent width $\sigma$ of the Gaussian packet is found from eq. (%
\ref{<n>Gauss}) in the form 
\begin{equation}
\sigma^2 = \sigma_0^2 + {\frac{\Omega^2 }{4 \sigma_0^2 \Delta^2}} (1 - \cos
\Delta t) \Big(1 + \cos(\Delta t - 2 \theta) \Big)\;.  \label{sigmat}
\end{equation}
Hence the variance of the position deviates from its initial value by an
oscillating term. The amplitude of this oscillation is governed by the ratio 
$(\Omega/2 \Delta \sigma_0)^2$. The initial width is restored whenever one
of the terms in brackets vanish. This happens twice during every Bloch
period, except when $\theta = \pi/2$, when these two instants coincide. This
combined breathing and oscillating behavior is illustrated in Figs. 2 and 3,
for various values of the relatice phase $\theta$. Notice that the
oscillation is always harmonic with the Bloch frequency $\Delta$. This is
due to the simple form of the dispersion relation for the case of
nearest-neighbor interaction. The time dependence of the variance is a
superposition of terms with frequencies $\Delta$ and $2 \Delta$.

\subsection{Zero external force}

In the absence of the external force, we can take the limit $\Delta
\rightarrow 0$ in the results of the previous subsections. In particular,
this gives $\phi = \beta = 0$, $\alpha(t) = \Omega t$. Then the evolution of
an initial Wannier state $|\Psi(0)\rangle = |0\rangle$ is given by 
\begin{equation}
|\Psi(t)\rangle = R |\psi(0)\rangle = \sum_n i^{-n} J_{n}(\Omega t)
|n\rangle\;,  \label{noforce}
\end{equation}
which shows that the free spreading of an initial Wannier state after a time 
$t$ gives Wannier populations equal to $p_n = |J_n(\Omega t)|^2$ \cite{VIS97}%
. The mean-square displacement increases linearly in time, as $\sigma =
\Omega t/\sqrt{2}$. This shows that the spreading is unbounded in the
absence of an external force. The self-propagator $p_0(t)$ decays to zero
for large times. The phase difference between neighboring sites is $\pm
\pi/2 $ at all times. For only two coupled wells, the coupling would give
rise to Rabi oscillations with frequency $\Omega$. Equation (\ref{noforce})
can be viewed as the generalization to the case of an infinite chain of
wells.

For a Gaussian wavepacket with initial width $\sigma_0$ and initial
quasimomentum determined by $\theta$, expressions (\ref{<n>t}) and (\ref
{sigmat}) take the form 
\begin{equation}
\langle n(t) \rangle = - \Omega t \sin \theta\;,\; \sigma^2 = \sigma_0^2 + {%
\frac{\Omega^2 t^2 }{8 \sigma_0^2}} (1 + \cos 2 \theta)\;.  \label{<n>sigmat}
\end{equation}
As one would expect in the absence of a uniform force, the group velocity
takes the constant value $v = - a \Omega \sin \theta$, which leads to
unbounded motion of the packet (except for $\theta = 0$ or $\pm \pi$).
Usually, the width increases indefinitely during he propagation. However,
for the special values $\theta = \pm \pi/2$ the width is constant, and the
packets propagates as a solitary wave. Notice that such a phase difference
between neighboring Wannier states arises spontaneously when a single
Wannier state spreads in the absence of a uniform force.

\section{Oscillating force}

Other situations of practical interest arise when the uniform force has an
oscillating component. Examples are the coupling between the states in the
Wannier-Stark ladder \cite{WIL96}, and dynamical localization for special
values of the amplitude-frequency ratio of the oscillation \cite{DUN86,BHA99}%
. The situation of an oscillating force is also decribed by the operator
description of Sec. \ref{evolution}. We give some results below.

\subsection{AC force only}

The situation of a harmonically oscillating uniform force can be expressed
as 
\begin{equation}
\Delta(t) = \delta \cos(\omega t)\;,  \label{oscforce}
\end{equation}
so that $\phi = (\delta/\omega) \sin (\omega t)$. Then according to (\ref
{alphabeta}) the parameters $\alpha$ and $\beta$ are specified by the
equalities 
\begin{equation}
\alpha e^{i\beta} = \Omega t J_0({\frac{\delta }{\omega}}) + \Omega \sum_{n
\neq 0} J_n({\frac{\delta }{\omega}}) {\frac{1 }{i n \omega}} \Big( %
e^{in\omega t} -1 \Big)\;,\;  \label{albetosc}
\end{equation}
where we used the expansion defining the ordinary Bessel functions, given in
section \ref{evolution}.

The first term in (\ref{albetosc}) increases linearly with time, whereas the
summation is bounded, and periodic in time with period $T = 2 \pi / \omega$.
The behavior of $\alpha$ and $\beta$ as defined by (\ref{albetosc}) is quite
complicated in general. However, for large times the value of $\alpha$, and
thereby the spreading of an initial Wannier state, is the same as in the
absence of the uniform force, with $\Omega$ replaced by the reduced
effective coupling $\Omega J_0(\delta/\omega)$. After one period $T$, the
values of the parameters become simple, and we find $\beta = \phi = 0$, $%
\alpha = \Omega T J_0(\delta / \omega)$. The evolution operator $U(T)$
during one period $T$ is simply given by the operator $R$ defined in (\ref
{Ut}), at these values of the parameters. The eigenstates of the evolution
operator $R = U(T)$ are simply the Bloch states $|k \rangle$. The
eigenvalues can be expressed as the $\exp(-i {\cal E}(k) T / \hbar)$, with 
\begin{equation}
{\cal E}(k) = \hbar \Omega J_0({\frac{\delta }{\omega}})  \label{quasiE}
\end{equation}
the corresponding values of the quasienergy, which are strictly speaking
only defined modulo $\hbar \omega$. The quasienergy bandwidth is reduced by
the factor $J_0(\delta / \omega)$, compared with the energy bandwidth in the
absence of the uniform force.

When the ratio $\delta/\omega$ of the amplitude and the frequency of the
oscillating force coincides with a zero of the Bessel function $J_0$, no
unbounded spreading occurs, and an initially localized state remains
localized at all times, with a periodically varying mean-square
displacement. The quasienergy bandwidth is reduced to zero in this case.
This effect of dynamical localization has been discussed before for
electrons in crystals \cite{DUN86}. A related effect of an effective
switch-off of atom-field coupling occurs for a two-level atom in a
frequency-modulated field when the ratio of the amplitude-frequency ratio of
the modulation equals a zero of the Bessel function $J_0$. This effect,
which leads to population trapping in a two-level atom, has recently been
discussed by Agarwal and Harshawardhan \cite{AGA94}.

\subsection{AC and DC force}

A constant uniform force creates Wannier-Stark states with equidistant
energy values. An additional oscillating force can induce transitions
between these states. Therefore, we consider the force specified by 
\begin{equation}
\Delta(t) = \Delta_0 + \delta \cos(\omega t)\;.  \label{cstoscforce}
\end{equation}
Then the value of the parameters $\phi$, $\alpha$ and $\beta$ are 
\begin{equation}
\phi(t) = \Delta_0 t + (\delta/\omega) \sin (\omega t)\;,\; \alpha
e^{i\beta} = \Omega \sum_n J_n({\frac{\delta }{\omega}}) {\frac{1 }{i
(\Delta_0 + n \omega)}} \Big( e^{i(\Delta_0 + n\omega) t} -1 \Big)\;.
\label{defparosct}
\end{equation}
In general, each term in the summation is bounded and periodic, but the
different periods can be incompatible. Moreover, whenever $\Delta_0 + n
\omega = 0$, the corresponding summand attains the unbounded form $\Omega t
J_n(\delta/\omega)$. At such a resonant value of $\Delta_0$, the spreading
of an initially localized state becomes unbounded, and the particle becomes
delocalized. This delocalization is suppressed again when the ratio $%
\delta/\omega$ is equal to a zero of the corresponding Bessel function $J_n$%
. This is a simplified version of the phenomenon of fractional Wannier-Stark
ladders, which has recently been observed and discussed \cite{MAD99,DIE00}.

The quasienergy values are again determined by the eigenstates of the
evolution operator $U(T)$ for one period of the oscillating force. This
operator is equal to the general operator $R$ defined in (\ref{Ut}), with
the parameters 
\begin{equation}
\alpha = 2 \Omega \sin (\Delta_0 T/2) \sum_n J_n({\frac{\delta }{\omega}}) {%
\frac{1 }{\Delta_0 + n \omega}}\;,\; \beta(T) = \Delta_0 T/2\;,\;\phi(T) =
\Delta_0 T \;.  \label{defparoscT}
\end{equation}
These expressions are correct whenever $\Delta_0 + n \omega$ is nonzero for
all values of $n$. Since these values of the parameters can be directly
mapped onto the values (\ref{defparcst}) specifying the evolution with a
constant uniform force, also the eigenvectors and corresponding
quasienergies are immediately found. The eigenvectors of $R$ can be
expressed as $|\psi_m \rangle = \sum_n |n\rangle c_n^{(m)}$, with the
expansion coefficients $c_n^{(m)} = J_{m - n}(\zeta)$. Here the argument $%
\zeta$ of the Bessel functions must be chosen as the sum 
\begin{equation}
\zeta = \Omega \sum_n J_n({\frac{\delta }{\omega}}) {\frac{1 }{\Delta_0 + n
\omega}}\;,  \label{eigenwanosc}
\end{equation}
which replaces the simple argument $\Omega / \Delta$ in eq. (\ref{eigenwan}%
). The eigenvalues of $R = U(T)$ are $\exp(-i {\cal E}_m T/\hbar)$, with the
discrete quasienergy values ${\cal E}_m = \hbar m \Delta_0$ (modulo $\hbar
\omega$).

In the resonant case that $\Delta_0 + n_0 \omega = 0$ for some integer $n_0$%
, one summand in the expression for $\alpha$ and $\beta$ is modified, as
indicated above. When $T = t$, only this modified summand is nonzero, and
the evolution operator $U(T) = R$ for one time period is characterized by
the values 
\begin{equation}
\alpha = \Omega T J_{n_0}\;,\;\beta = 0\;,\;\phi = - 2 \pi n_0\;.
\label{defparspec}
\end{equation}
The eigenvectors of $R$ are the Bloch states $|k \rangle $, and the
corresponding quasienergy values are 
\begin{equation}
{\cal E}(k) = \hbar \Omega J_{n_0} ({\frac{\delta }{\omega}})\;.
\label{quasiEres}
\end{equation}

\section{Discussion and conclusions}

We have analyzed the Wannier-Stark system, which is characterized by the
Hamiltonian (\ref{Hdisc}), in terms of the operators $B_{\pm}$ and $B_0$.
The present interest in this model arises from the dynamics of atoms in a
periodic optical potential, with an additionally applied uniform external
force. We adopted the tight-binding limit, which implied nearest-neighbor
interaction only. This gives rise to an explicit simple dispersion relation
between energy and quasimomentum, which makes the model exactly solvable.
From the commutation properties of the basic operators we obtain eq. (\ref
{Ut}) for the evolution operator for an arbitrary time dependence of the
uniform force, where the three parameters are defined in (\ref{phi}) and (%
\ref{alphabeta}). As shown in Secs. \ref{heis} and \ref{specinit}, the
parameter $\phi$ determines the shift in the value of the quasimomentum,
whereas $\alpha$ and $\beta$ determine the evolution of the average position
and the width of a wavepacket. A particle starting in a single Wannier state
has a uniform distribution over the quasimomentum, and cannot change its
average position, wheras the width of its wavepacket is simply measured by $%
\alpha$. On the other hand, even when only two neighboring states are
populated initially, the wavepacket can display an appreciable motion. In
Sec. \ref{Gaussian} it is demonstrated that an initially Gaussian packet
remains Gaussian at all times. This remains true when the initial state has
a non-zero expectation value of the quasimomentum, which is described as an
initial phase difference between neighboring Wannier states.

These results, which are valid for a uniform force with an arbitrary time
dependence, unify and extend earlier results obtained for a constant or an
oscillating uniform force. A constant force induces Bloch oscillations of a
wavepacket, and we obtain a simple expression for the amplitude of the
oscillation and for the time dependence of the width of a wavepacket. For an
oscillating force, the operator method shows that the quasienergy bands can
be evaluated directly in terms of the value of the parameter $\alpha$ after
one oscillation period. This produces an exactly solvable model for
dynamical localization and fractional Wannier-Stark ladders. In general, by
selecting a proper time-dependence of the force or of the coupling between
wells, thereby realize the desired values of the parameters $\alpha$, $\beta$
and $\phi$, we can coherently control the width and the position of a
wavepacket, as well as the phase difference between neighboring sites.

\acknowledgments

This work is part of the research program of the ``Stichting voor
Fundamenteel Onderzoek der Materie'' (FOM).

\begin{figure}[tbp]
\caption{Upper part: plot of the breathing population distribution for an
initial Wannier state $|0 \rangle$. Lower part: plot of the oscillating
population distribution, for two initial superposition of Wannier states $|0
\rangle$ and $| 1 \rangle$, and two different values of the relative phase $%
\theta$. Both plots are evaluated for $\Omega / \Delta = 6$. Shaded
distributions hold after half a Bloch period $t = \pi / \Delta$.}
\label{fig1}
\end{figure}

\begin{figure}[tbp]
\caption{Periodic behavior of the width and the average position of a
Gaussian wavepacket for various initial values of the phase difference $%
\theta$ between neighboring states. Initial value of the width is $\sigma_0
= 4$, and $\Omega / \Delta = 50$.}
\label{fig2}
\end{figure}

\begin{figure}[tbp]
\caption{Bloch oscillation and corresponding breathing behavior of a
Gaussian wavepacket in a constant uniform force. Values of $\sigma_0$, and $%
\Omega$ and $\Delta$ as in Fig. \ref{fig2}. Upper part: $\theta = 0$. Lower
part: $\theta = \pi/2$.}
\label{fig3}
\end{figure}

\end{document}